\def\aa{{A\&A}}

\def\aj{{AJ}}
\def\annrev{{ARA\&A}}
\def\apj{{ApJ}}
\def\apjs{{ApJS}}

\def\mnras{{MNRAS}}
\def\nat{{Nature}}

\documentclass[cup5b]{caps}
\usepackage{graphicx}
\usepackage{amssymb}
\usepackage{ociwsymp1}
\HeadText{A. Burkert and T. Naab}

\def\plotone#1{\centering \leavevmode
\includegraphics[width=.95\columnwidth]{#1}}

\begin{document}

\pagenumbering{arabic}

\author[]{ANDREAS BURKERT$^1$ and THORSTEN NAAB$^2$ \\
(1) Max-Planck-Institut f\"ur Astronomie, K\"onigstuhl 17, D-69117 Heidelberg, 
Germany \\
(2) Institute of Astronomy, Madingley Road, Cambridge CB3 0HA, UK}

\chapter{The Formation of Spheroidal \\ Stellar Systems}

\begin{abstract}
We summarize current models of the formation of spheroidal stellar systems. 
Whereas globular clusters form in an efficient mode of star formation inside 
turbulent molecular clouds, the origin of galactic spheroids, that is bulges, 
dwarf ellipticals, and giant ellipticals, is directly coupled with structure 
formation and merging of structures in the Universe. Disks are the fundamental 
building blocks of galaxies and the progenitors of galactic spheroids. The 
origin of the various types of spheroids and their global properties can be 
understood as a result of disk heating by external perturbations, internal 
disk instabilities, or minor and major mergers.
\end{abstract}

\section{The Realm of Spheroids}

Spheroids exist in the Universe with a wide range in masses and length scales. 
Probably the most simple, classical examples of stellar spheroids are globular star
clusters with masses in the range of $10^4\, M_{\odot}$ to $10^6 \,M_{\odot}$ and half-mass radii of order
2--10 pc (Harris 1996). These almost spherical systems appear to be stable and very long lived. Although the metallicities 
of different clusters in the Milky Way vary from [Fe/H] $\approx\, -2.5$ to solar or even larger, the strikingly narrow
iron abundance spreads of stars within individual clusters (Kraft 1979) and their small age spread
indicate that each cluster consists of
only one stellar generation that formed on a short time scale from chemically homogenized gas.
Peebles \& Dicke (1968) proposed that globular clusters are the first objects that formed
in the Universe. More recent models assume that globulars formed at the same time as their host galaxies 
(Fall \& Rees 1985; Vietri \& Pesce 1995).
As giant molecular clouds have similar masses and radii, they are considered to be the primary sites of cluster formation.
Unfortunately, the formation of stars and the condensation of 
molecular clouds into dense, massive star clusters is still not well understood up to now 
(for a recent review see Lada \& Lada 2003). 
Klessen \& Burkert (2000, 2001) and Bate, Bonnell, \& Bromm (2002; see also Clarke, this volume) investigated numerically
the gravitational collapse of a turbulent cloud. Their models showed that the 
stabilizing turbulent motion of the molecular gas is dissipated 
on a short dynamical time scale, resulting in
collapse and star formation. These models, however, neglected energetic feedback processes, which are known to
play a crucial role in regulating and terminating star formation.
In order to form a gravitationally bound, dense stellar cluster,
high local star formation efficiencies of order $\eta_{\rm sf} \approx 50\%$ are required 
(Brown, Burkert, \& Truran 1991, 1995; Geyer \& Burkert 2001). This is in 
contradiction with observations, which indicate that the fraction of molecular cloud material that
turns into stars is typically of order $\eta_{\rm sf} \leq 10\%$ due to gas ionization  by the 
UV field of newly formed high-mass stars (Myers et al. 1986; Williams \& McKee 1997; Koo 1999).
Ashman \& Zepf (1992) argued that globular clusters can form efficiently 
in interacting galaxies (Schweizer 1999), which indicates that peculiar, galactic
non-equilibrium environments might enhance the star formation efficiency in molecular clouds.
Under these conditions, 
supersonic cloud-cloud collisions or cloud implosions induced by an increase of the external 
gas pressure could destabilize a whole cloud complex, triggering global collapse and efficient
star formation.

\begin{figure}
\plotone{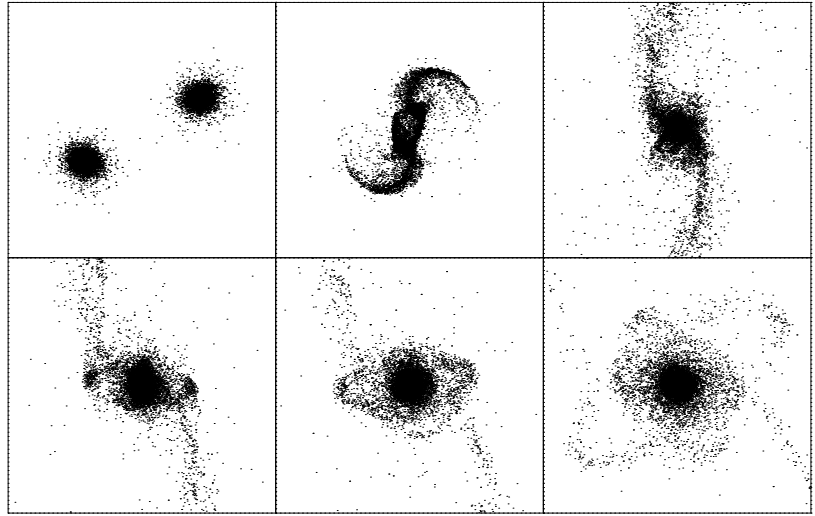}
\caption{Simulation of an equal-mass spiral galaxy merger. Each box has a size
of 210 kpc. The upper-left box shows the initial condition. The other boxes
show snapshots of the evolution at $4.6 \times 10^8$, $8 \times 10^8$,
$1.3 \times 10^9$, $1.9 \times 10^9$, and $2.6 \times 10^9$ yr. Dark matter 
deficient dwarf spheroidals might form through gravitational instabilities 
inside tidal arms. The merger remnant is surrounded by rings and shells that 
provide long-term signatures of its merging history.}
\label{figure1}
\end{figure}

In contrast to globular clusters, the origin and structure of galactic spheroids (dwarf spheroidals, dwarf ellipticals, giant ellipticals, and bulges) seem to be strongly coupled with the hierarchical merging history of substructures 
in the Universe. Within the popular cold dark matter (CDM) cosmogony, the visible components of galaxies arise from gas infall
into dark matter halos, followed by star formation. Disks are envisioned to form as a result of smooth gas accretion from
the intergalactic medium (e.g., Katz \& Gunn 1991; Navarro \& White 1994; Steinmetz \& M\"uller 1994). Spheroids result
from processes that heat and destroy stellar disks. Low-mass dwarf spirals
are particularly sensitive to stirring and harassment by the cumulative tidal interactions of high-speed galaxy encounters 
in galactic clusters (Moore et al. 1996; Moore, Lake, \& Katz 1998), leading in the end to dwarf ellipticals.
Massive spiral galaxies, on the other hand, can be destroyed
by major mergers (Fig. 1.1) and transform into giant ellipticals and bulges (Toomre 1974; Kauffmann, Charlot, \& White 1996).

Recently, numerical simulations have shown that the formation of disks and sphe\-roids might be even more complex.
High-resolution cosmological simulations of galaxy evolution 
including star formation and feedback processes (Steinmetz \& Navarro 2002) as well as semi-analytical models (Khochfar \& Burkert 
2003) indicate that the galaxies change their morphological type frequently.
For example, spheroids could form by an early merger of low-mass disks and
later on rebuild new disks by smooth gas accretion. These bulge-disk systems could merge
again, forming an even larger spheroid. Within the framework of this scenario galactic bulges represent early spheroids that 
have grown a new, surrounding disk component. It is, however, not clear up to now whether all bulges necessarily formed that way
(e.g., Wyse, Gilmore, \& Franx 1997).
Wyse \& Gilmore (1992) argued that the specific angular momentum distribution of the Milky Way's bulge is very similar
to that of the stellar halo and very different from that of the disk. This would suggest that the bulge was built
up by dissipative inflow (Gnedin, Norman, \& Ostriker 2000) of gas that was lost from star-forming regions and substructures 
in the  Galactic halo, suggestive of a monolithic collapse scenario (Eggen, Lynden-Bell, \& Sandage 1962).
Yet another possibility are
disk instabilities (Athanassoula 2002), which lead to barlike structures that later on
transform into bulges through a buckling instability 
(Combes et al. 1990; Pfenniger \& Norman 1990; Norman, Sellwood, \& Hasan 1996; Noguchi 2000).
Balcells et al. (2003) report a lack of bulges with $r^{1/4}$ surface density profiles,
expected in the merging scenario, 
favoring the secondary process. Ellis, Abraham, \& Dickinson 
(2001) find that intermediate-redshift bulges 
are bluer than their elliptical counterparts, which indicates that bulges are younger than ellipticals, 
in contradiction with the bottom-up 
structure formation scenario of the CDM model.
It is likely that some bulges formed by disk instabilities and others by early mergers.
In this case, two bulge populations should exist, with different kinematic and photometric properties.

Within the cosmological CDM scenario, galactic spheroids are surrounded by dark matter halos. An exception might
be tidal tail galaxies. Distinct gaseous and stellar clumps are frequently found in tidal tails of interacting galaxies 
(Schweizer 1978; Mirabel, Lutz, \& Maza 1991). Barnes \& Hernquist (1992) used numerical simulations to 
demonstrate that these self-gravitating systems consist preferentially of gas and stars and form frequently
in the thin, expanding tails of merging galaxies. In contrast to structures that form by cosmological merging,
the  dark matter fraction in tidal tail galaxies is negligibly small. 
The dwarf spheroidals orbiting the Milky Way might represent such a 
population of dark matter deficient tidal tail systems (Irwin \& Hatzidimitriou 1995; Klessen \& Kroupa 1998).

The formation of galactic spheroids as a result of discrete, violent perturbations of galactic disks is supported
by the observation that galaxy populations vary strongly with the galaxy density in clusters.
It has been recognized early that most early-type systems are found in clusters 
(e.g., Hubble \& Humason 1931).  Detailed observations by Dressler (1980) suggested a well-defined
relationship between the local density in clusters and galaxy type (see also Whitmore \& Gilmore 1991). Postman \& Geller (1984)
extended the study of this morphology-density relation to poorer groups of galaxies and defined a single morphology-density relation
that is valid over 6 orders of magnitude in density. Melnick \& Sargent (1977) found a relation between the morphological type of
individual galaxies and their distance from the cluster center. It is still a matter of debate whether this morphology-radius
relation follows from the morphology-density relation, or vice versa. Whitmore, Gilmore, \& Jones (1993) argued on the basis
of Dressler's data that the distance from the cluster center is the more fundamental parameter. This conclusion
is supported by the study of Sanrom\'a \& Salvador-Sol\'e (1990) who showed that the radial variations in cluster
properties are preserved independent of substructure. 

{\it Hubble Space Telescope}\ images of clusters at intermediate redshifts have confirmed that morphological transformations occur frequently
in clusters. Dressler et al. (1997) and Couch et al. (1998) found an abnormally high proportion of spiral and irregular types at
redshifts $z \approx 0.5$ and an increase of the fraction of S0 galaxies toward the present time.
These observations are in agreement with cosmological models that predict that galaxy mergers lead
to ellipticals and S0s, and that in dense, rich clusters no subsequent formation of a new disk component is possible
(Kauffmann, White, \& Guiderdoni 1993; Baugh, Cole, \& Frenk 1996; Kauffmann 1996). 
Okamoto \& Nagashima (2001) combined semi-analytical methods
with cosmological $N$-body simulations to study the formation and evolution of cluster galaxies. Their models can reproduce the 
morphology-density relation for elliptical galaxies. However they also predict a clear separation between bulge-dominated
and disk-dominated galaxy types in clusters. Mixed types like S0 galaxies should be rare, which is not in agreement with the
the observations.

\section{Rotating Spheroids}

\begin{figure}
\plotone{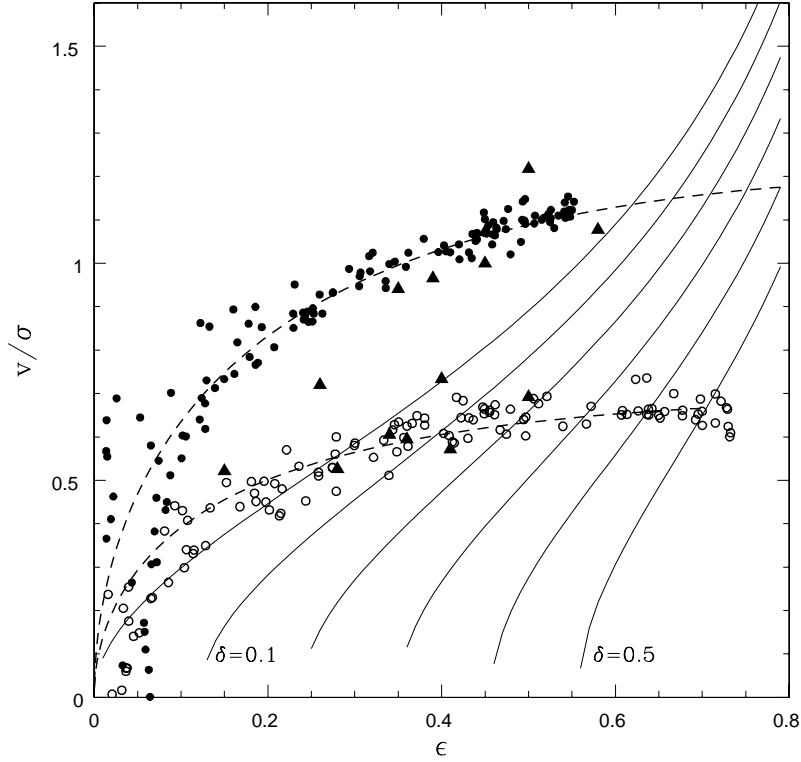}
\caption{The ratio of line-of-sight rotational velocity to line-of-sight 
velocity dispersion as a function of ellipticity for disky ellipticals 
(triangles) and two collisionless merger remnants of disk galaxies (filled and
open circles), viewed with different projection angles.  The solid lines show
theoretical predictions of anisotropic stellar systems with given anisotropy 
$\delta$.  The dashed curves show inclination effects for a system with 
$(\epsilon,\delta)=(0.78,0.4)$ and $(\epsilon,\delta)=(0.7,0.5)$.}
\label{sample-figure}
\end{figure}

Stellar equilibrium systems exist in two basic configurations: rotationally supported disks and
pressure-supported spheroids. 
Disks are stabilized by the balance between centrifugal forces and gravity. Their
radial surface density distribution is determined primarily by the specific 
angular momentum distribution of the stellar system and the
shape of the gravitational potential well or the total mass distribution.
The velocity dispersion $\sigma$ of stellar disks is, by definition, small compared to their rotation
$\upsilon_{\rm rot}$. It therefore does not affect their radial density profiles or rotation curves, 
while still regulating their vertical thickness.

Disks are called dynamically cold because of their small random velocities
$\sigma \ll \upsilon_{\rm rot}$. Spheroids are, in contrast, dynamically hot stellar systems with $\sigma \geq \upsilon_{\rm rot}$. 
Even in these systems angular momentum and rotation can still play an important role.
The difference between disks and ellipticals is therefore not necessarily a result of differences in 
the specific angular momentum distribution but rather due to differences in the stellar
velocity dispersion. Stellar disks, for example, can easily be converted into spheroids through internal instabilities or
external perturbations that increase the particles' vertical velocity dispersion,
even if their angular momentum distribution remains unchanged. This scenario is very attractive in explaining
the origin of dwarf ellipticals, which have exponential surface brightness profiles, reminiscent of a disk progenitor.
A process that could convert exponential disks into spheroids is tidal interaction in clusters 
(Moore et al. 1998). Galactic harassment, however, would not
reduce significantly the rotational velocity, in contrast with recent observations by
Geha, Guhathakurta. \& van~der~Marel (2002).

Spheroids could either be flattened by rotation or by an anisotropic velocity distribution.
Violent processes that break up disks and lead to ellipticals should in general result in 
anisotropic systems. However, it has been argued that especially lower-mass, disky ellipticals
are rotationally flattened and isotropic systems (Bender 1988a).  
If the equidensity surfaces of an oblate spheroid with
ellipticity $\epsilon$ are all similar, the ratio of line-of-sight
rotational velocity $\upsilon$ to line-of-sight velocity dispersion $\sigma$ is (Binney \& Tremaine 1987)

\begin{equation}
\frac{\upsilon^2}{\sigma^2} = 0.5 (1-\delta) \frac{\frac{\arcsin \epsilon}{\epsilon}-\sqrt{1-\epsilon^2}}
{\sqrt{1-\epsilon^2}-(1-\epsilon^2)\frac{\arcsin \epsilon}{\epsilon}} -1
\end{equation}

\noindent where the anisotropy parameter $\delta=1-\Pi_{zz}/\Pi_{xx}$
measures the deviation from isotropy and $\Pi_{ii}$ is the random kinetic energy tensor component in the i'th direction. 
The solid lines in Figure 1.2 show $\upsilon/\sigma$ versus $\epsilon$ for various 
values of $\delta$. For $\epsilon > 0.2$, inclination (dashed curves) mainly decreases the ellipticity, with no significant
change in $\upsilon/\sigma$.  The triangles in Figure 1.2 show observed lower-mass disky ellipticals. They appear to be
isotropic, with $\delta < 0.2$. However, some objects, especially those with $\upsilon/\sigma \approx 0.5$, could also represent inclined
anisotropic ellipticals, with intrinsic anisotropies of $\delta = 0.5$ and high ellipticities ($\epsilon \approx 0.7$). 
If seen edge-on, these systems would be interpreted as S0 galaxies and therefore would not be classified
as disky ellipticals. The open and filled circles show two merger remnants from numerical simulations
with mass ratios 3:1 and different initial disk orientations
(Burkert, Naab, \& Binney 2003, in preparation). Each point represents a different projection angle and 
follows the theoretical dependence
of $\upsilon/\sigma$ and $\epsilon$ on the inclination angle. We find that mergers of initially aligned disks result in ellipticals that
are indeed intrinsically isotropic (filled circles) and fast rotating, with $\upsilon/\sigma = 1$.
Misaligned disks, however, form
ellipticals that are anisotropic (open circles) with $\delta=0.5$ and $\upsilon/\sigma = 0.5$. These objects could still
appear isotropic due to inclination effects.

\section{Stellar Equilibrium Systems}

Any stellar dynamical system is completely specified  by its phase space distribution function $f(\vec{x},\vec{\upsilon},t)$, 
which determines the number of stars that at time $t$ have positions $\vec{x}$ in a small volume $dx^3$ and 
velocities $\vec{\upsilon}$ in the small range $d\upsilon^3$. In collisionless systems the flow of points in  the
6-dimensional phase space resembles an incompressible fluid and is determined by the Vlasov equation $df/dt=0$.
In equilibrium $f$ must be a steady-state solution ($\partial f/\partial t=0$) of the Vlasov equation, and the Jeans theorem holds, 
which says that $f$ depends on the phase space coordinates only through integrals of motion. In the case of  spherical
symmetry with an isotropic velocity dispersion, $f$ is only a function of the energy: $f=f(E=\upsilon^2/2-\Phi)$, where
$\Phi$ is the gravitational potential.
Obviously there exist an infinite number of equilibrium distribution functions, and stellar spheroids 
could have a large variety of density distributions. This is not observed, however. 
Galaxies can be subdivided just into two major groups with respect to their density profiles: giant ellipticals  and
dwarf ellipticals. Giant ellipticals are characterized by de~Vaucouleurs profiles (de~Vaucouleurs 1948; Kormendy 1977),  dwarfs
by exponential profiles. The exponential profiles might be reminiscent of exponential progenitor disks.
The origin of the de~Vaucouleurs profile and the observed regularity in giant ellipticals is
more obscure and still not completely understood.

Internal secular evolution due to two-body relaxation (e.g., Lynden-Bell \& Wood 1968) could efficiently 
erase the information about the initial state, leading to universal structures. This is likely in the case of 
globular clusters with lifetimes that are large compared to their internal relaxation time scale. 
The situation, however, is different for galaxies, which have 
two-body relaxation time scales that  by far exceed their age.
Hernquist (1990) presented an analytical density distribution $\rho_{\rm H}(r)$ that closely 
matches the de~Vaucouleurs law:

\begin{equation}
\rho_{\rm H}(r)=\frac{M}{2 \pi} \frac{a}{r} \frac{1}{(r+a)^3}, 
\end{equation}
\noindent where $M$ is the total mass and $a$ is a scale length.

The velocity dispersion profile $\sigma(r)$ in the inner region of the Hernquist spheroid is given by

\begin{equation}
\sigma^2 \sim r \ \ln \left( \frac{a}{r} \right)
\end{equation}

\noindent and is characterized by a kinematically cold, power-law density core with
a velocity dispersion that decreases toward the center and a density that diverges for $r \rightarrow 0$. 
Numerical simulations of
galaxy mergers confirm that kinematically cold cores form as predicted by Equation 1.3.
Binney (1982) calculated the fractional energy distribution $N(E)$ that would be required for a stellar systems
to follow the  $r^{1/4}$ law. He found the interesting result that $N(E)$  is well described by a Boltzmann law

\begin{equation}
N(E)=N_0 \exp(\beta E), 
\end{equation}

\noindent where $\beta=-2 r_e/GM$ represents a negative temperature. Although such an energy distribution is also found
in numerical simulations (Spergel \& Hernquist 1992) there does not yet exist any 
analytical theory that could explain its origin.  

The origin of universal $r^{1/4}$ profiles might require a phase of strong violent relaxation of the stellar system.
Lynden-Bell (1967) noted that strong fluctuations of the
gravitational potential during this relaxation phase would change the specific energy distribution of
stellar systems and might eventually lead to a universal relaxed state that is independent of the initial
conditions. Subsequently, orbital phase mixing will drive the systems toward equilibrium on a time scale
of order 2--3 dynamical time scales. Simulations of violently collapsing
collisionless stellar systems (van~Albada 1982) lead to equilibrium states that were in rough
agreement with a de~Vaucouleurs profile. A universal state, however, is only achieved
if the initial density distribution is very concentrated, as otherwise phase space
constraints affect the relaxation and final structure of the inner region (Burkert 1990; Hozumi, Burkert, \& Fujiwara 2000).
Spergel \& Hernquist (1992) adopted a different approach and proposed that violent relaxation can be described by  numerous random 
orbital perturbations that occur preferentially at perigalacticon. In this case, the probability of a particle being scattered into
a given state would be proportional to the phase space accessible at perigalacticon, resulting in an exponential energy distribution.

\begin{figure}
\hspace{-1.5cm}
\includegraphics[width=.95\columnwidth,angle=-90]{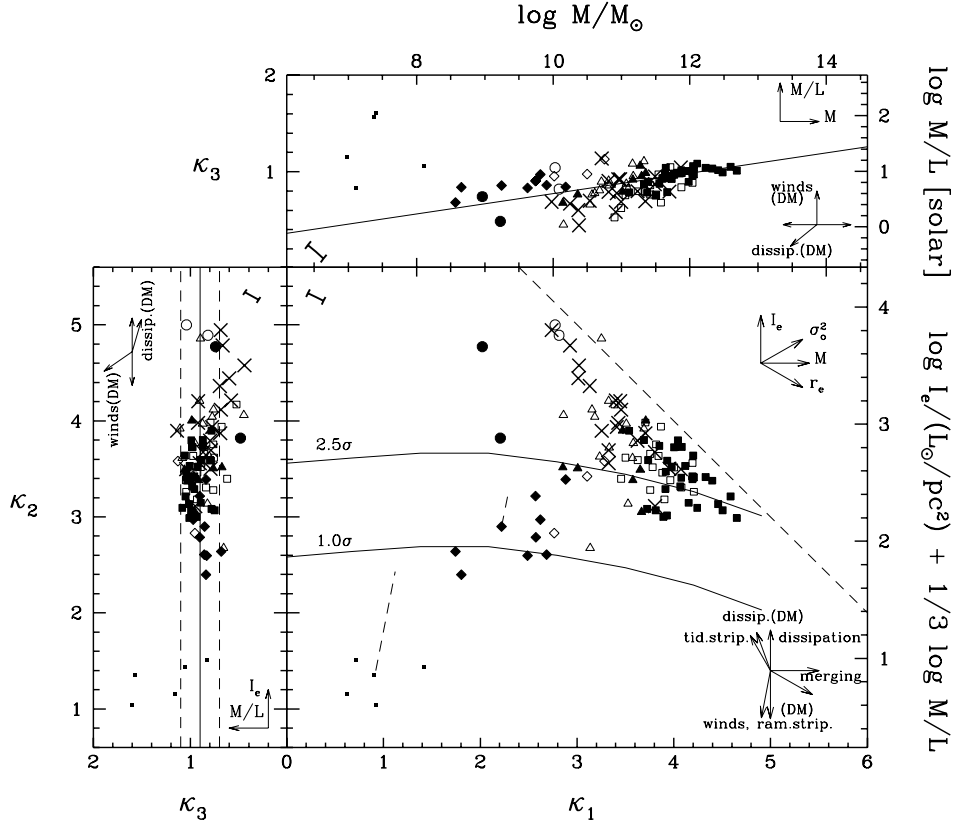}
\caption{This figure, adopted from Bender, Burstein, \& Faber (1997), shows 
the distribution of all types of dynamically hot galaxies in $\kappa$-space. 
Large squares denote giant ellipticals ($M_T<-20.5$ mag); triangles show 
ellipticals of intermediate luminosity (--20.5 mag $< M_T <$ --18.5 mag). 
Circles and diamonds denote compact ellipticals and dwarf galaxies, 
respectively. Open symbols are rotationally flattened galaxies, while filled 
symbols are anisotropic objects. Bulges are represented by crosses. The five 
small filled squares at low $\kappa_1$ values denote local dwarf spheroidals.
The set of arrows indicates how dissipation with and without dark matter, 
tidal stripping, ram pressure stripping, or merging  would move the objects in 
$\kappa$ space. The curved lines marked $1.0 \sigma$ and $2.5 \sigma$ indicate 
]the range of $\kappa_1$ versus $\kappa_2$ values expected from a CDM density
fluctuation spectrum neglecting dissipation.}
\label{sample-figure}
\end{figure}

\section{Fundamental Plane Relations}

Stellar systems are characterized by three global physical parameters: central velocity dispersion $\sigma_0$, effective radius
$r_e$, and effective surface brightness $\mu_e$, or, in physical units, $\log I_e = -0.4(\mu_e-27)$. With $L \sim I_er_e^2$
and assuming virial equilibrium ($M \sim \sigma_0^2 r_e$) Bender, Burstein, \& Faber (1992) introduced an orthogonal coordinate
system in the 3-space of the observable parameters $\log \sigma_0^2$, $\log r_e$ and $\log I_e$:

\begin{eqnarray}
\kappa_1 \equiv (\log \sigma_0^2 + \log r_{\mathbf{e}})/\sqrt{2},\\
\kappa_2 \equiv (\log \sigma_0^2 +2 \log I_{\mathbf{e}} -\log
r_{\mathbf{e}})/\sqrt{6},\\
\kappa_3 \equiv(\log \sigma_0^2 - \log I_{\mathbf{e}} -\log
r_{\mathbf{e}})/\sqrt{3}.
\end{eqnarray}

If we define the luminosity $L$ and the mass $M$ of a galaxy as $L=c_1 I_{\mathbf{e}}r_{\mathbf{e}}^2$ and $M=c_2
\sigma_0^2r_{\mathbf{e}}$, as given by the virial theorem, with $c_1$ and
$c_2$ being structure constants, the effective radius can be written as $r_{\mathbf{e}}=
(c_1/c_2)(M/L)^{-1}\sigma_0^2 I_{\mathbf{e}}^{-1}$. Then $\kappa_1$
is proportional to $\log M$, $\kappa_2$ is proportional to $\log (M/L) I_{\mathbf{e}}^3$, and 
$\kappa_3$ is proportional to $\log (M/L)$.

Figure 1.3 shows the distribution of elliptical galaxies 
and bulges in $\kappa$-space. The $\kappa_1 - \kappa_3$ projection shows the plane edge-on.
Its tilt is independent of the environment (J{\o}rgensen, Franx, \& Kjaergaard 1996) and does in 
general also exist for S0s and dwarf ellipticals (Nieto et al. 1990). In addition to the optical, a fundamental plane
is also found in the infrared, but with a slightly different
slope (Mobasher et al. 1999), and probably in the X-ray regime (Fukugita \& Peebles 1999).
The origin of the slope is not well understood up to now. It probably corresponds
to variations in the internal structure and to changes in metallicity and age, which seem to correlate well
with galaxy mass.

The edge-on view of the fundamental plane can be thought of as a consequence of the virial theorem, independent
of initial conditions.
The face-on view ($\kappa_1 - \kappa_2$ projection), on the other hand, provides important information about
the formation of spheroids. In this plane dwarf ellipticals and giant ellipticals divide into two 
orthogonal sequences (see also Kormendy 1985; Binggeli \& Cameron 1991). 
Whereas giant ellipticals and bulges with total blue luminosities brighter than $M_{B_T} \approx -18$ mag  and 
stellar masses $M_* > 10^{10}\, M_{\odot}$ are characterized by high surface densities that decrease 
systematically with 
increasing mass, dwarf ellipticals with $M_{B_T} \geq -18$ mag are diffuse and have surface densities that increase 
with mass or luminosity. Dissipationless collapse in a CDM Universe would produce structures that lie within
the thin solid lines denoted $1.0 \sigma$ and $2.5 \sigma$. Energy dissipation moves galaxies toward larger $\kappa_2$ values. 
Obviously, low-mass giant ellipticals and bulges experienced a large amount of dissipation, leading to high
surface densities, compared to the expected dissipationless values. Giant ellipticals, on the other hand, might have
formed in gas-poor stellar mergers, which are preferentially dissipationless. 
The sequence of dwarf ellipticals that runs almost perpendicular to giant ellipticals indicates that
these systems might have strongly been affected by wind-driven mass loss (Larson 1974; Arimoto \& Yoshii 1986, 1987; 
Dekel \& Silk 1986; Vader 1986;  Matteucci \& Tornamb\`{e} 1987; Martinelli, Matteucci, \& Colafrancesco 2000), which decreased both
$\kappa_2$ and $\kappa_1$. The galactic wind model can also explain the observed color-magnitude relation
(Faber 1973; Bower, Lucey, \& Ellis 1992), according to which the
integrated colors of dwarf ellipticals become progressively bluer toward fainter luminosities.
Gas loss would terminate the epoch of star formation progressively later in more massive ellipticals
with deeper potential wells. The stellar populations in brighter galaxies should therefore be
more enhanced in heavy elements and would appear redder.
Bender et al. (1997) argued, however, that progressively larger amounts of mass loss,
starting from a single progenitor galaxy with $\kappa_1 \approx 3.5$ and $\kappa_2 \approx 2.6$ 
cannot explain the dwarf sequence, which in this case should be much steeper. Dwarf galaxies instead
had to form from different progenitors with different initial densities and probably also different
amounts of mass loss. It is still not clear up to now why a large range of possible progenitors and the
expected strong variations in star formation and galactic mass loss histories
should lead to dwarf ellipticals that populate such
a narrow one-dimensional sequence in $\kappa$-space.

The dichotomy between dwarf and giant ellipticals is clearly visible when 
investigating their global or central properties.
The situation seems to be different when one considers the shape of their light profiles, 
where the transition appears to be more continuous.
Most bright dEs have an inner luminosity excess above the exponential surface brightness profile that is characteristic
for low-luminosity dwarfs (Binggeli \& Cameron 1991). The profiles of these nucleated dwarfs resemble closely the characteristic
$r^{1/4}$ profiles of giant ellipticals. This observed continuity motivated Young \& Currie (1994) and subsequently
Jerjen \& Binggeli (1997) and Binggeli \& Jerjen (1998) to fit S\'{e}rsic (1968) profiles 

\begin{equation}
I(r) = I_0 e^{-(r/r_0)^{n}}
\end{equation}

\noindent to their sample of early-type dwarf and giant galaxies (see also Caon, Capaccioli, \& D'Onofrio 1993). 
They found that the S\'ersic index $n$ 
and the S\'{e}rsic parameters $I_0$ and $r_0$ vary smoothly with luminosity, indicating that all ellipticals
can be reunited into one sequence. The exception are compact ellipticals 
(Faber 1973; Burkert 1994), which are a rare and special kind of
ellipticals with shapes like giants but luminosities like dwarfs. Up to now it is not clear why
all ellipticals have surface brightness profiles that vary smoothly with luminosity while, at the same time,
their global parameters and also their central parameters (Kormendy 1985) show a clear dichotomy between
giant and dwarf ellipticals.

\section{The Formation of Elliptical Galaxies}

Elliptical galaxies have long been thought to be simple spheroidal
dynamically relaxed stellar systems that  follow a
universal de~Vaucouleurs $r^{1/4}$ law (de~Vaucouleurs 1948) and are 
classified only by their ellipticity.  
The traditional formation mechanism for giant ellipticals that
would naturally result in a homogeneous family of galaxies is the 
``monolithic collapse'' model. It was motivated by the idea
that the oldest stars of the spheroidal halo component of the
Galaxy formed during a short period of radial collapse of gas
(Eggen et al. 1962). In this case,
ellipticals could have formed very early as soon as
a finite over-dense region of gas and dark matter decoupled from the expansion of the Universe
and collapsed. If during the protogalactic collapse phase star formation was very efficient,
a coeval spheroidal stellar system could have formed (Partridge \& Peebles 1967; Larson 1969, 1974; Searle, Sargent, \& Bagnuolo 1973) 
before the gas dissipated its kinetic and potential energy and settled
into the equatorial plane, forming a disk galaxy.  
A possible test of this assumption is the
redshift evolution of the zero point of the fundamental plane, which 
is a very sensitive indicator of the age of a stellar population (van~Dokkum \& Franx 1996). It evolves
very slowly, especially for massive elliptical galaxies, indicating a formation redshift of their stars of $z \geq 3$
(Bender et al. 1998; van~Dokkum et al. 1998).
This scenario would be in agreement with the monolithic collapse picture.

An alternative scenario, proposed by Toomre \& Toomre (1972), is
that elliptical galaxies formed via a morphological transformation
induced by binary mergers of disk galaxies. During the merging phase
the stellar disks experienced a phase of 
violent relaxation due to the strong tidal interactions,
resulting in a spheroidal merger remnant. 
The merging scenario has been tested by observations of rich
clusters at intermediate and low redshifts.
There is growing evidence that the abundance of spiral galaxies in
clusters indeed decreases from a redshift of $z= 0.8$ to $z=0$
(Dressler et al. 1997; Couch et al. 1998; van~Dokkum et al. 2000). A similar trend is observed for the relative numbers
of star-forming and post-starburst galaxies 
(Butcher-Oemler effect) (Butcher \& Oemler 1978, 1984; Postman, Lubin, \& Oke 1998; Poggianti et al. 1999). 
At the same time, the early-type fraction increases from 40\% to 80\% between $z=1$ and $z=0$
(van~Dokkum \& Franx 2001). Semi-analytical models of galaxy formation within
the hierarchical merging scenario by  Kauffmann (1996) and Kauffmann \& Charlot (1998)
are also consistent with a low formation redshift for early-type galaxies, which seems
to be in contradiction with the ages of their stellar populations.
Van~Dokkum \& Franx (2001) showed that this problem can be solved if the
progenitors of present-day ellipticals are not classified
as ellipticals at high redshift. In this case, the apparent luminosity and color evolution
would look similar to a single age stellar population that formed at very high
redshift, independent of the true star formation history.

\subsection{Boxy and Disky Ellipticals}

Further insight into the formation history of ellipticals comes from
detailed observations of nearby galaxies, which
can be subdivided into two groups with respect to
their structural properties (Bender 1988a; Bender, D\"obereiner, \& M\"ollenhoff 1988; Kormendy \& Bender 1996). 
Faint  giant ellipticals are isotropic rotators with small minor 
axis rotation and  disky deviations of their isophotal contours from
perfect ellipses.
Their diskiness might be due to a faint secondary disk component that contributes up to
30\% to the total light in these galaxies. 
Disky ellipticals also have power-law inner density
profiles (Lauer et al. 1995; Faber et al. 1997) and show little or no radio and X-ray emission
(Bender et al. 1989).  Bright giant elliptical galaxies with 
$L_B \ge 10^{11}\,L_odot$, 
on the other hand, exhibit nearly elliptical or box-shaped isophotes and show flat cores. 
Their kinematics are generally more complex than those of disky objects. They rotate slowly, are
supported by anisotropic velocity dispersions and have
a large amount of minor axis rotation.
Boxy galaxies have smaller values of $n$ than disky galaxies. 
Occasionally, they have
kinematically distinct cores (Bender 1988b; Franx \& Illingworth 1988; Jedrzejewski \& Schechter 1988), 
which are metal enhanced, indicating that gas infall and subsequent star
formation must have played some role during their formation 
(Bender \& Surma 1992; Davies, Sadler, \& Peletier 1993). Boxy ellipticals also show stronger
radio emission than average and have high X-ray luminosities, 
consistent with emission from hot gaseous halos (Beuing et al. 1999). 

The distinct physical properties of disky and boxy elliptical galaxies
demonstrates that the two types of ellipticals could have experienced different
formation histories. It has been argued by Kormendy \& Bender (1996)
and Faber et al. (1997) that the high surface densities (see Fig. 1.3), the
secondary disk components, and the central
power-law density cusps of disky ellipticals result from  substantial gas dissipation
during the merging of gas-rich progenitors.  Disky ellipticals seem to continue the
Hubble sequence from S0s to higher bulge-to-disk ratios.
Boxy ellipticals, on the other hand, might have formed by dissipationless mergers between collisionless stellar disks 
or other ellipticals (Naab \& Burkert 2000; Khochfar \& Burkert 2003).

\subsection{Merger Simulations} 

Merger simulations of disk galaxies provide the best access to a direct
comparison with observations of individual galaxies. The stellar
content of a galaxy is represented by particles that can be
analyzed with respect to their photometric and kinematic properties
in the same way as an observed galaxy. It has generally been assumed that the
progenitors of ellipticals galaxies are disk galaxies. That this assumption is 
questionable has been demonstrated by Khochfar \& Burkert (2003).
Their semi-analytical models show that most massive ellipticals actually formed 
by mixed (elliptical-spiral) or early-type mergers.

Negroponte \& White (1983), Barnes (1988), and Hernquist (1992) performed the 
first fully self-consistent merger models of two equal-mass stellar
disks embedded in dark matter  halos. The remnants were slowly
rotating, pressure-supported, anisotropic, and generally
followed an $r^{1/4}$ surface density profile in the outer
parts. However, due to phase space limitations (Carlberg 1986) the surface brightness profiles in the
inner regions were flatter than observed. To solve this problem
a massive central bulge component had to be included in
the progenitors (Hernquist 1993a). In this case, the
progenitors resembled already early-type galaxies. It seems to be unlikely that
all merger progenitors of ellipticals contained a
massive central bulge component. On the other hand,
these simulations already emphasized that global
properties of equal-mass merger remnants resemble those of ordinary, 
slowly rotating massive elliptical galaxies. 

Additional evidence for the merger scenario are tidal tails and shells that are observed
in the outer parts of ellipticals and are found to be a natural result of disk mergers
(Hernquist \& Spergel 1992). In addition, the formation of kinematically decoupled
subsystems in merger simulations that include gas strongly support 
the merger scenario (Hernquist \& Barnes 1991). Note, however, that Harsoula \& Voglis (1998) proposed an alternative
scenario where kinematically distinct subsystems can form directly from
an early cosmological collapse without any major mergers thereafter. 

More detailed investigations of the isophotal shapes of equal-mass merger
remnants have shown that the same remnant can appear either disky or
boxy when viewed from different directions (Hernquist 1993b). 
This result is puzzling
since most  boxy ellipticals are radio and X-ray luminous, in contrast to disky
ellipticals. As radio and X-ray properties should not depend on projection effects, 
the isophotes should not change with viewing angle.

\begin{figure}
\plotone{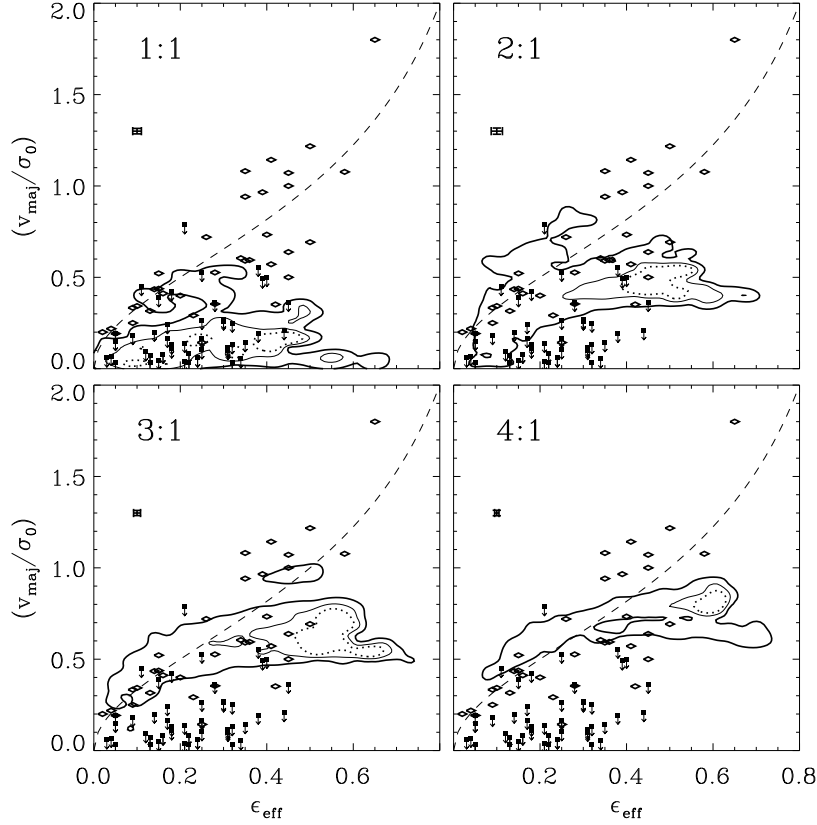}
\caption{ Rotational velocity over velocity dispersion versus characteristic
ellipticity for mergers with various mass ratios. Values for observed ellipticals are overplotted. 
The dashed line shows the theoretically predicted correlation for an oblate isotropic rotator.}
\label{figure4}
\end{figure}

In contrast to anisotropic, equal-mass mergers, 
mergers with a mass ratio of 3:1 lead to remnants that are
flattened and fast rotating (Bendo \& Barnes 2000).
Naab, Burkert, \& Hernquist (1999) analyzed the photometric and kinematic
properties of a typical 3:1 merger remnant and
compared the results to observational data of disky elliptical
galaxies. They found an excellent agreement and proposed that fast-rotating 
disky elliptical galaxies can originate from pure  
collisionless 3:1 mergers, as opposed to slowly rotating, pressure-supported
ellipticals, which might form from equal-mass mergers of disk galaxies.  
Burkert \& Naab (2003) and Naab \& Burkert (2003) analyzed a large number of
high-resolution, statistically unbiased mergers with mass ratios of 1:1,
2:1, 3:1, and 4:1.  They concluded that the
dichotomy of giant ellipticals can be understood as a sequence of mass ratios of
disk-disk mergers (Fig. 1.4). Equal-mass mergers produce anisotropic and slowly 
rotating remnants with a large amount of minor axis rotation. A subset
of initial disk orientations result in purely boxy ellipticals. Only if 
the initial spins of the disks are aligned will the remnant appear 
isotropic and disky or boxy depending on the orientation.  
In contrast, 3:1 and 4:1 mergers form a more homogeneous group of remnants. 
They have preferentially disky isophotes, and are fast rotating with
small minor axis rotation, independent of the assumed projection. 
2:1 mergers have intermediate properties, with boxy or disky isophotes
depending on the projection and the orbital geometry of the merger. 

The influence of gas on the global structure of elliptical galaxies is not well understood. 
Observations indicate that some giant ellipticals contain a significant amount of gas that is distributed in an extended
disklike component (Oosterloo et al. 2002; Young 2002). Such an extended disk
naturally forms in gas-rich, fast-rotating, 3:1 merger remnants (Naab \& Burkert 2001).
Even in 1:1 mergers the 
remaining gas in the outer parts of the remnant has high enough angular momentum to form extended 
gas disks as it falls back (Barnes 2002).  On the other hand, the simulations of equal-mass mergers also 
indicate that half of the gas is driven to the center of the remnant, producing a peak in surface density
that is not observed (Mihos \& Hernquist 1994; Barnes \& Hernquist 1996).

The presence of gas in merger simulations influences the stellar structure of the remnants. 
Even if star formation is neglected, 
stars in remnants of gas-rich mergers are less likely to be on box orbits than their 
collisionless counterparts (Barnes \& Hernquist 1996), leading to a better agreement with observations
of stellar line-of-sight velocity distributions (Bender, Saglia, 
\& Gerhard 1994; Naab \& Burkert 2001).
The influence of star formation on merger remnants has theoretically
been addressed in detail by Bekki \& Shioya (1997), Bekki (1998), and more
recently by Springel (2000).  They found that the rapidity of gas
consumption can affect the isophotal shapes. Secular star formation, 
however, leads to final density profiles that deviate
significantly from the observed $r^{1/4}$ profiles in radial
regimes where all ellipticals show almost perfect de~Vaucouleurs
laws (Burkert 1993). As star formation is likely to occur in all disk galaxy mergers this result represents
a serious  problem for the merger scenario.

\section{Conclusions}
Within the framework of cosmological hierarchical structure formation, galactic disks represent the 
fundamental building blocks where most of the stars form. Tidal encounters and galaxy mergers heat and
destroy these disks, resulting in kinematically hot stellar systems. Galaxy harassment in clusters can
preserve the disk structure while increasing the random kinetic energy of the stars perpendicular to the disk.
In this case, exponential dwarf ellipticals would form. Galaxy mergers represent more violent processes 
that lead to strong violent relaxation, erasing the information about the initial state and resulting in a 
de~Vaucouleur's profile as seen in giant elliptical galaxies. 

Detailed observations of the kinematic and geometric properties of spheroids, coupled with sophisticated
high-resolution simulations,  have led to major progress in understanding the origin of these systems.
However, many problems still exist and need to be investigated in detail.

Violent relaxation and the origin of the $r^{1/4}$ law is still not understood up to now.
Observations of nonrotating, exponential dwarf ellipticals are in contradiction with the harassment scenario. In
addition, there exists no theory that can predict why the scale length of dwarf ellipticals is on average
in the range of 0.5--1 kpc, independent of luminosity.
More observations are required to test the theoretical predictions of two different bulge populations, one
with exponential profiles, resulting from disk instabilities, and the other with de~Vaucouleur's profiles,
resulting from an early, violent merger phase of the protogalaxy.
It is also not clear whether the dichotomy of giant ellipticals into disk and boxy objects is preferentially due
to variations in the mass ratio of the merger components. Another possibility is an additional gaseous component
that settled into an equatorial disk inside the spheroid where it turned into stars. In this case, gas dynamics and dissipation will
have affected the structure preferentially in disky ellipticals, which might explain their
high surface densities compared to massive, boxy ellipticals that formed preferentially by dissipationless mergers.
More simulations including gas dynamics and star formation are required in order to test this scenario.
The origin of the most luminous giant ellipticals is currently not understood at all. These objects are
much more massive than disk galaxies and therefore could not have formed by major disk mergers. In addition, their metallicities are
supersolar and higher than the stellar populations of disk galaxies. Luminous, giant ellipticals probably formed by
multiple mergers within dense groups of galaxies followed by an 
efficient phase of star formation and metal enrichment. Whether this scenario can also explain
their large ages and their location in the low-density region of 
the fundamental plane needs to be explored in greater detail.

\vspace{0.3cm}
{\bf Acknowledgements}.
Andreas Burkert would like to thank Luis Ho for the invitation to a very stimulating and pleasant conference.

\end{document}